&latex209
\documentclass[12pt]{iopart}
\usepackage{epsfig}
\usepackage{graphicx}
\newcommand{\bra}{\langle}
\newcommand{\ket}{\rangle}

\newcommand{\bmu}{\mbox{\boldmath $\mu$}}

\newcommand{\be}{\begin{equation}}
\newcommand{\ee}{\end{equation}}
\newcommand{\bd}{\begin{displaymath}}
\newcommand{\ed}{\end{displaymath}}

\def\vecm{\mbox{\boldmath $m$}}

\def\veczeta{\mbox{\boldmath $\zeta$}}
\def\vecxi{\mbox{\boldmath $\xi$}}
\def\vecmu{\mbox{\boldmath $\mu$}}
\def\matA{\mbox{\boldmath $A$}}

\def\bs{\mbox{\boldmath $s$}}

\begin{document}
\title[
Study of an associative memory model with multiple stable states
]
{Retrieval Properties of Hopfield and Correlated Attractors
in an Associative Memory Model
}
\author{
Tatsuya UEZU, Aya HIRANO and Masato OKADA$^{1,2}$}
\address{  
Graduate School of Humanities and Sciences,
 Nara Women's University, Nara 630-8506, Japan\\
$^1$ Laboratory for Mathematical Neuroscience,\\
RIKEN Brain Science Institute, Saitama 351-0198, Japan\\
$^2$ ``Intelligent Cooperation and Control'', PRESTO, JST,\\
c/o RIKEN BSI, Saitama 351-0198, Japan
}
\begin{abstract}
We examine a previouly introduced attractor neural network model 
that explains the persistent activities of neurons
in the anterior ventral temporal cortex of the brain.
In this model, the coexistence of several attractors
including correlated attractors was reported
in the cases of finite and infinite loading.
In this paper, by means of a statistical mechanical method,
we study the statics and  dynamics of the model 
in both  finite and extensive loading, 
mainly focusing on the retrieval properties of
the Hopfield and  correlated attractors.
In the extensive loading case, we derive
 the evolution equations by the dynamical replica theory. 
We found several characteristic temporal behaviours,  
both in the finite and extensive loading cases.
The theoretical results were confirmed by numerical simulations.
\end{abstract}

 \ead{uezu@ki-rin.phys.nara-wu.ac.jp}

\newfont{\msamfnt}{msam10}
\newcommand{\gsim}{\mbox{$\\maketitle$ \msamfnt\symbol{'046}
$\\maketitle$}}
\newcommand{\llsim}{\mbox{$\\maketitle$ \msamfnt\symbol{'056}
$\\maketitle$}}
\def\veco{\mbox{\boldmath $w^o$}}
\def\vecx{\mbox{\boldmath $x$}}
\def\vecw{\mbox{\boldmath $w$}}
\def\veceta{\mbox{\boldmath $\eta$}}

\section{Introducion}
The attractor neural network model has been considered 
a plausible model for memory processing in the brain.
Many experimental findings implying its existence have been reported.
Among them, Miyashita and his group reported the persistent activities of
neurons in the anterior ventral 
temporal (AVT) cortex.\cite{Miyashita88,MiyashitaChan88}
They trained monkeys to recognize and match a set of 100 colored
fractal patterns in the Delayed Match to Sample (DMS) paradigm.
Miyashita and Chang reported that the neurons in the AVT are highly
selective toward a few of the 100 colored fractal patterns,
and that the activities of these neurons
 persist for 16 seconds after the removal of the
presented stimulus.\cite{MiyashitaChan88}
Miyashita studied development of this selectivity,
and found that serial positions number of stimuli during learning 
are converted into the spatial correlation 
of the neural activities.\cite{Miyashita88}

To explain these experiments, Griniasty et al. proposed a model 
for an attractor neural network.\cite{Griniastyetal93}
By analysing equilibrium states of this model in the 
finite loading case,
 they found not only stored patterns (Hopfield attractors)
 but also patterns which have overlap 
with stored patterns (correlated attractors). 
For example, when the number of patterns 
is 13,   correlated attractors are expressed
by $\vecm=\frac{1}{2^7}^t(77,51,13,3,1,0,0,0,0,1,3,13,51)$
and its cyclic rotations, whereas  Hopfield attractors
are expressed by  $\vecm=^t(1,0,\cdots,0)$
and its cyclic rotations. 
Here, 
$\vecm$ is the overlap vector 
$\vecm=^t(m_1,m_2,\cdots,m_{c})$, of which component 
$m_{\mu}$ is the overlap between the state of  neurons 
and the $\mu$-th pattern. $t$ denotes transposition. See eq. (\ref{eq:6}). 
Griniasty et al. found that the correlation  
between correlated attractors dicreases as the separation 
in the sequence of the patterns to which the attractors belong
increases.
These results support the work done by Miyashita.

In recent years, we have studied the statics and dynamics of 
the model introduced by Griniasty et al.
 in the finite loading case.
As for the statics, we investigated equilibrium states and found the 
 coexistence of several attractors such
as Hopfield attractors, correlated attractors, and mixed states
of  several patterns in several parameter regions.  
The coexistence of attractors was also reported
by Shiino et al.\cite{Shiino97,ShiinoYamano98} 
and by Fukai et al.\cite{Fukaietal99}
As for the dynamics, we  investigated the transient behaviour
by performing numerical calculations of the overlap dynamics
derived theoretically and by numerical simulations.
In particular, we studied what happens when
a Hopfield attractor and a correlated attractor 
coexist and when 
 a Hopfield attractor does not exist and a correlated attractor
does exist.
In the former case, we studied the basin of attraction
and in the latter case, we found that 
 the trajectory initially approaches the state where the Hopfield
 attractor existed,
but finally it tends toward the correlated attractor.
We found that the results by numerical calculations
of the overlap dynamics and the results
by numerical simulations agree qualitatively.

The coexistence of multiple attractors and the dynamical behaviour
obtained in the  finite loading case 
are significant.  However, 
the situation of the finite loading
seems to be unrealistic.
 Thus, it is necessary to investigate whether the results
in the  finite loading case
are obtained when applied practically, i.e. in the
 extensive loading case.

In this paper, we study the  model introduced by Griniasty et al.
 by the statistical mechanical method,
mainly focusing on the extensive loading case.
In the extensive loading case, as to the study of the equilibrium states,
Cugliandolo et al. derived the saddle point equations(S.P.E.)
for order parameters 
using the replica method.\cite{CugliandoloTsodyks94}
They identified not only the Hopfield attractors but also
the correlated attractors.
We have solved the saddle point equations numerically by scanning
parameters  and  found that there are parameter regions
where  several attractors coexist. 
The coexistence of multiple attractors was found by Shiino et al.
in the present model as well.\cite{Shiino97,ShiinoYamano98}
  As for the dynamical behaviour, we have 
derived evolution equations for order parameters
by using the dynamical replica 
theory by Laughton and Coolen.\cite{LaughtonCoolen95}
As in the finite loading case,
 we have studied the dynamical behaviour in a coexistence region 
of a Hopfield attractor and a correlated attractor,
and also in   a   region where a Hopfield attractor does not exist
but a correlated attractor does exist.
From these studies, we have obtained  qualitatively similar
results to those in the finite loading case.

In the next section, we explain the present model.
Then, we summarize the theory and results for  the finite loading case
in \S3. In \S4, the extensive loading case is examined.
The summary and discussion
are given in \S5.

\section{Model}

Let us explain the model.  The instantaneous state of each neuron
 is expressed by $s_i$ which takes $\pm 1$, where $i$
labels the neuron($i=1,\cdots,N$),  and the time evolution is given by
\begin{equation}
 s_i(t+1) = {\rm sign}(h_i(t)), \label{eq:amit1}
\end{equation}
where 
\begin{equation}
 h_i(t) = \sum_{j (\ne i)} J_{ij}s_j(t),\label{eq:amit2}
\end{equation}
and  $J_{ij}$ is the strength of the synaptic connection
from the $j$-th neuron to the $i$-th neuron.
The stochastic dynamics is also taken into consideration
by introducing temperature $T$,
that is, the probability that  $s_i(t+1)$ takes $\pm 1$ is given by
\begin{equation}
{\rm Prob}[s_i(t+1)] =\frac{1 \pm \tanh(\beta h_i(t))}{2},\label{eq:amit3}
\end{equation}
where $\beta = 1/T$. 
In the present model,
 the synaptic weight  $J_{ij}$ is defined as
\begin{eqnarray}
J_{ij} &= & \frac{1}{N} \sum_{\mu =1}^{c}
(  \xi _i ^{\mu} \xi _j ^{\mu} +a  \xi _i ^{\mu} \xi _j ^{\mu -1}
+a \xi _i ^{\mu} \xi _j ^{\mu +1})
=   \frac{1}{N} \sum_{\mu,\nu}
\xi_i ^{\mu} A_{\mu \nu} \xi_j ^{\nu}
\;\; \mbox{ for $i \ne j$}, \label{eq:amit4}\\
&&J_{ii}= 0,\;\; \xi ^0 _i \equiv \xi ^{c} _i,\;\;
\xi ^{c +1} _i \equiv \xi^1 _i,
\nonumber
\end{eqnarray}
where $\xi _i ^{\mu}$ represents the value of $i$-th neuron
for $\mu$-th pattern $\vecxi ^{\mu} \equiv \;^t(\xi^{\mu}_1, \cdots,
\xi_N ^{\mu})$ and it takes value +1 or -1
with a probability of 1/2. $c$ is the total number of patterns. 
$\matA$ is a $c \times c$
 matrix defined as
\begin{eqnarray}
\matA & \equiv & \{ A_{\mu \nu}\} =
\left(
\begin{array}{rrrrrrr}
   1 &  a      & 0 &\cdots    & 0  & a \\
   a &  1      & a & 0 & \cdots   & 0\\
   0 &  a      & 1 & a & 0        & 0 \\
   0 &  0      & a & 1 & a        & 0 \\
\vdots & \ddots & \ddots& \ddots  & \ddots &\vdots\\
   a & 0 &  \cdots &  0  & a & 1
\end{array}
\right). \label{eq:matrix}
\end{eqnarray}
The order parameter is the overlap vector
$\vecm=^t(m_1,m_2,\cdots,m_{c})$, of which component 
$m_{\mu}$ is the overlap between the state of  neurons 
$\bs = ^t(s_1, \cdots, s_N)$ and 
the $\mu$-th pattern $\vecxi ^{\mu} $, that is,
\begin{equation}
m_{\mu}=\frac{1}{N}\sum _{i=1}^N s_i \xi _i^{\mu}.
\label{eq:6}
\end{equation}

\section{Finite Loading Case}

In this section, we study the finite loading case of the model
described by eqs.(1)-(4).
That is, we consider what happens when $N\gg 1$ and $\alpha=\frac{c}{N}\ll 1$.

\subsection{Theory}
To study the equilibrium state, we consider the following 
Hamiltonian $H$.
\begin{eqnarray}
H & = & - \frac{1}{2}\sum_{i \ne j}J_{ij}s_i s_j\\
&=&  - \frac{1}{2N}\sum_{i \ne j} \sum_{\mu,\nu}
\xi_i ^{\mu} A_{\mu \nu} \xi_j ^{\nu} s_i s_j.
\end{eqnarray}
We calculate the free energy of the system
\[
 f=-\frac{1}{\beta N}\ln Z, \;\;Z={\rm Tr}_{\bs}e^{-\beta H},
\]
where Tr$_{\bs}$ denotes the summation with respect to
$s_i (i=1, \cdots, N )$.
By using the saddle point method, we obtain\cite{Herzetal}
\begin{eqnarray}
f & =& - \frac{\beta}{2}\sum_{\mu, \nu}m_{\mu}A_{\mu \nu}m_{\nu}
-i\sum_{\mu=1}^c m_{\mu}\hat{m}_{\mu}
-[ \ln \{2 \cosh(-i \sum_{\mu=1}^c \hat{m}_{\mu}\xi^{\mu})\}]_{\vecxi},
\end{eqnarray}
where $\vecxi=^t(\xi^1,\xi^2,\cdots,\xi^c)$ and
$[A]_{\vecxi}$ is the average over 
$\vecxi$, that is $[A]_{\vecxi}=\frac{1}{2^c}
\sum_{\{\xi ^\mu = \pm 1\}}$A.
Then, the S.P.E. becomes
\begin{eqnarray}
\hat{m}_{\mu} & =& i \beta \sum_{\nu=1}^c A_{\mu \nu}m_{\nu},\\
m_{\mu} & =& [\xi^{\mu} \tanh( \beta
\sum_{\nu,\nu '=1}^c \xi ^{\nu} A_{\nu \nu'}m_{\nu '}) ]_{\vecxi}.
\label{eq:alpha0spe}
\end{eqnarray}
The physical meaning of $m_{\mu}$ is the overlap between the
equilibrium state and the $\mu$ -th pattern,
\begin{equation}
 m_{\mu}=\frac{1}{N}\sum_{i=1}^N \xi _i ^{\mu} \bra s_i \ket,
\end{equation}
where $\bra s_i \ket$ is the thermal average of the $i$-th neuron.

Next, we derive the evolution equation of the system.
Let us consider the probability ${\cal P}_t(\vecm)$
that at time $t$ the state $\bs$ has the overlap $m_{\mu}$
with $\mu$-th pattern $\vecxi ^{\mu}$ for $\mu=1,\cdots,c$
\begin{eqnarray}
{\cal P}_t(\vecm)&=& {\rm Tr}_{\bs} p_t(\bs)\prod_{\mu=1}^c
\delta(m_{\mu}-m_{\mu}(\bs)),\\
m_{\mu}(\bs)&=&\frac{1}{N}\sum_{i=1}^N \xi _i ^{\mu} s_i,
\end{eqnarray}
where $p_t(\bs)$ is the probability that the system takes the state 
$\bs$ at time $t$.
We assume that the transition probability $w_{i}(\bs)$
 from the state $\bs = (s_1, \cdots, s_i,\cdots, s_N)$
to the state $F_i \bs = (s_1, \cdots, - s_i,\cdots, s_N)$
 takes the following form
\begin{equation}
w_{i}(\bs)
=\frac{1-s_i\tanh\{\beta h_{i}(\bs)\}}{2},
\end{equation}
where $F_{i}$ is the flip operator of the $i$-th neuron
and $h_i(\bs)$ is expressed as
\begin{eqnarray}
h_{i}(\bs)= \sum_{j(\ne i)}^N J_{ij}s_j \simeq
\sum_{\mu,\nu \leq c}
\xi_{i}^{\mu}A_{\mu \nu} m_{\nu}(\bs).
\end{eqnarray}
The master equation for $p_{t}(\bs)$ is given by
\begin{eqnarray}
\frac{d}{dt}p_{t}(\bs)
& = & \sum _{i=1}^N
\{  w_{i}(F_{i}\bs) p_{t}(F_{i}\bs)
      - w_{i}(\bs)  p_{t}(\bs) \}.
\end{eqnarray}
Then, using asynchronous dynamics, we obtain  the
evolution equations for the overlap $\vecm$
as
\begin{eqnarray}
\frac{d}{dt}\vecm &=& 
-\vecm +[ \vecxi
      \tanh( \beta \; ^t \vecxi \matA \vecm) ]_{\vecxi}.
\label{eq:alpha0ev1}
\end{eqnarray}
The equation for the stationary state of eq. (\ref{eq:alpha0ev1})
agrees to the eq. (\ref{eq:alpha0spe}) for the equilibrium state.

\subsection{Results}
$c$ is finite and we put $c=13$ 
 as in ref.\cite{CugliandoloTsodyks94},
 becasuse the correlated attractors obtained by Amit et al.
have non-zero values up to their fifth nearest neighbours.\\

\noindent
{\bf The equilibrium state}

We solved the S.P.E.(\ref{eq:alpha0spe}) numerically and
found the coexistence of  several attractors such
as the Hopfield attractors, correlated attractors, and mixed states
of several patterns in several parameter regions.  
For example, we show the result for
 $a=0.4$ in fig. 1.  The solid curve denotes the Hopfield attractor.
It exists up to $T \simeq 0.1$.  The dashed curves denote
a correlated attractor.  For this attractor,
there is the following symmetry 
\[
 m_2=m_{13},\;\; m_3=m_{12},\cdots, m_7=m_{8}.
\]
This attractor exists up to $T \simeq 0.25$.
It has been considered that correlated attractors exist only
for $a>0.5$.\cite{CugliandoloTsodyks94}
However, we found that they exist even for $a<0.5$.
As is seen from the figure, up to $T \simeq 0.1$, both
the Hopfield attractor and the correlated attractor exist.
Further, we found that the mixed states of 
three patterns $\vecxi^{\mu}$,
$\vecxi^{\mu +1}$ and $\vecxi^{\mu +2}$ 
(which are not drawn in fig. 1) exist up to $T \simeq 0.05$
and the mixed state of 13 patterns exists up to $T \simeq 1.7$.\\

\noindent
{\bf Dynamics}

First, we investigate the case where several attractors coexist.
For example, for $a=0.4$ and $T=0.04$, 
a Hopfield attractor, a correlated attractor, mixed states of
three patterns and a mixed state of 13 patterns coexist.
For this parameter, we investigate the basin of attraction of
these attractors by using overlap dynamics (eq. (\ref{eq:alpha0ev1}))
and numerical simulations.
In the numerical integrations of eq. (\ref{eq:alpha0ev1}),
we set 
the initial condition as
\begin{equation}
m_{\mu}(t=0) = m_0 \delta_{\mu,1}, \;\; \mu=1, \cdots, c. \label{eq:inicon}
\end{equation}
On the other hand, in the numerical simulations,
we generate an initial state $\bs(0)$ according to
the following probability,
\begin{equation}
{\rm Prob}[s_i=\pm 1] = \frac{1 \pm m_0 \xi _i ^1}{2}.
\label{eq:prob}
\end{equation}
When $N \rightarrow \infty$, the overlap $m_{\mu}$
 between $\bs(0)$ and $\vecxi ^{\mu}$
satisfies the relation (\ref{eq:inicon}).\\
In fig. 2, we show the results by numerical integrations
and those by numerical simulations.  In fig. 3, we compare the
results by numerical integrations and those by
numerical simulations in more detail.
As is shown in fig. 3(a), in the numerical integrations,
when the initial overlap with the pattern 1, $m_0$, is
0.15, the trajectory tends toward the correlated attractor
and the trajectory with $m_0=0.16$ tends toward the Hopfield attractor.
That is,  the boundary between the basin of attraction
 for the Hopfield attractor and that for the correlated attractor, $m_0 ^c$,
is between 0.15 and 0.16.
We show the results of the simulations in fig. 3(b).
When $m_0$ is 0.16, the trajectory tends toward the correlated attractor and
when  $m_0$ is 0.17, the trajectory tends toward the Hopfield attractor.
From these results, we note that even if the trajectories finally tend toward
the correlated attractor, they initially approach the Hopfield 
attractor.  

Next, we investigate the case
 where a Hopfield attractor does not exist but
a correlated attractor exists, which is shown in fig. 4.  As
the figures show, both 
in the numerical integrations of  eq. (\ref{eq:alpha0ev1}) and
numerical simulations, we found that for $m_0<1$,
 the trajectory initially approaches the state where the Hopfield
 attractor existed, but finally it tends toward the correlated attractor.\par
In the next section, we study the extensive loading case to see
whether the characteristic feature in the finite loading case such as
the coexistence of multiple attractors and characteristic temporal behaviour
continues to exist or not.

\section{Extensive Loading Case}
\subsection{Theroy}
As an example of an extensive loading case, we consider the following 
 synaptic weight  $J_{ij}$,
\begin{eqnarray*}
\hspace*{-2cm}
J_{ij} &= & \frac{1}{N}\{ \sum_{\mu =1}^{c}
(  \xi _i ^{\mu} \xi _j ^{\mu} +a  \xi _i ^{\mu} \xi _j ^{\mu -1}
+a \xi _i ^{\mu} \xi _j ^{\mu +1})
+ \sum_{\mu =c +1}^{p}  \eta _i ^{\mu} \eta _j ^{\mu}
\}
\;\; \mbox{ for $i \ne j$},\\
\hspace*{-2cm}
&&J_{ii}= 0,\;\; \xi_i ^0 \equiv \xi_i ^{c},\;\; 
\xi_i ^{c +1} \equiv \xi_i ^1.
\end{eqnarray*}
$c$ is the number of condensed patterns and we set $c=13$. 
$p$ is the total number of patterns and we consider
the case that $\alpha =\frac{p}{N}$ is finite.
We assume that $\xi_i ^{\mu}$ and $\eta_i ^{\mu}$ take +1
or -1 with a probability of 1/2.
The overlap $m_{\mu}(\bs)$ between the state of neurons
$\bs$ and the $\mu$-th pattern $\vecxi ^{\mu}$ is defined as
\begin{eqnarray*}
 m_{\mu}(\bs) &=& \frac{1}{N}\sum_{i=1} ^N s_i \xi ^{\mu} _i
\;\; \mbox{ for $\mu=1, \cdots, c$}\\
&=& \frac{1}{N}\sum_{i=1} ^N s_i \eta ^{\mu} _i
\;\; \mbox{ for $\mu=c+1, \cdots, p$}.
\end{eqnarray*}
 The cross-talk noise 
$z_{i}(\bs)$ is defined as
\begin{eqnarray*}
z_{i}(\bs)
&\equiv & {\sum}_{\mu > c }\eta _{i}^{\mu}m_{\mu}(\bs).
\end{eqnarray*}
As order parameters, we define  $\vecm=^t(m_1,m_2,\cdots,m_{c})$ 
 and  $r=\frac{1}{\alpha}\sum_{\mu=c+1}^p (m_{\mu}(\bs))^2$ where
$\alpha r$ is the variance of the cross-talk noise.\\

\noindent
{\bf The equilibrium states}

By using the replica method,
the free energy $f$ for the replica symmetric(RS)
solution\cite{CugliandoloTsodyks94} is given as
\begin{eqnarray}
 f& = & \frac{1}{2} {}^t\vecm \matA \vecm + \frac{\alpha}{2\beta}
  \left(
      \ln(1-\beta+\beta q)
      -\frac{\beta q}{1-\beta+\beta q}
  \right)
 +\frac{\alpha \beta}{2}r(1-q)\nonumber\\
&&-T [ \int Dz  \ln \{ 2 \cosh \beta(\sqrt{\alpha
 r}z+ ^t \veczeta \matA \vecm)\}] _{\veczeta},
\end{eqnarray}
where $\veczeta= ^t(\zeta^1, \cdots, \zeta ^{c})$
 and $[ A ] _{\veczeta}$ is the average over $\veczeta$,
that is ，$\frac{1}{2^c} \Sigma_{\{ \zeta ^{\mu}= \pm 1\}} A$.
The S.P.E. for  $\vecm$, $q$ and $r$ are given by
\begin{eqnarray}
\vecm &=& [\veczeta \int Dz \tanh \beta (^t \veczeta \matA \vecm +
\sqrt{\alpha r} z)] _{\veczeta}, \\\label{eq:sigmapemeq}
q &=& [ \int Dz  \tanh ^2 \beta (^t \veczeta \matA \vecm +
\sqrt{\alpha r} z)] _{\veczeta},\\
r&=& \frac{q}{(1-c)^2},
\end{eqnarray}
where $c \equiv  \beta (1-q)$,  $Dz = dz e^{-x^2/2}/\sqrt{2\pi}$ 
and $\matA$ is the $c \times c $ matrix which is
 defined in eq. (\ref{eq:matrix}).
For $T=0$, these equations become
\begin{eqnarray}
c& = & \sqrt{ \frac{2}{\pi \alpha r}}
[ \exp\{-(\frac{^t \veczeta \matA \vecm }{\sqrt{2\alpha r}})^2\}
 ] _{\veczeta},\label{eq:spe1}\\
\vecm &=& -2 [  \veczeta H(\frac{^t \veczeta \matA \vecm }
{\sqrt{\alpha r}})
] _{\veczeta},\label{eq:spe2}\\
r&=&\frac{1}{(1-c)^2},\label{eq:spe3}
\end{eqnarray}
where $H(x)= \int _x ^{\infty}\frac{dt}{\sqrt{2 \pi}}
e^{-t^2/2}$.\\

\noindent
{\bf Dynamics}

As for the dynamics, we derive the evolution equations for $(\vecm, r)$ 
using the dynamical replica theory(DRT).\cite{LaughtonCoolen95}
We assume that the same transition probability $w_{i}(\bs)$ as 
in the finite loading case, 
but in the present case,  $h_i(\bs)$ is expressed as
\begin{eqnarray}
h_{i}(\bs)=
\sum_{\mu,\nu \leq c}
\xi_{i}^{\mu}A_{\mu \nu} m_{\nu}(\bs)
+z_{i}(\bs).
\end{eqnarray}
The master equation for $p_{t}(\bs)$ is the same as before.
Using asynchronous dynamics, we obtain  the
evolution equations for the overlap $\vecm$ and the
variance of the cross-talk noise $\alpha r(\bs)$,
assuming the self-averaging in the limit of 
$N \rightarrow \infty$.
Then, we obtain
\begin{eqnarray}
\frac{d}{dt}\vecm &=& -\vecm + \int dz
              [ { \cal D}_{\veczeta}[z]\veczeta
      \tanh \beta(^t \veczeta \matA \vecm+z)
              ]_{\veczeta},\label{eq:ev1}\\
\frac{d}{dt}r&=& -r+1 + \frac{1}{\alpha}
                \int dz
                [ {\cal D}_{\veczeta}[z]z
                \tanh \beta(^t \veczeta \matA \vecm+z)
              ]_{\veczeta},\label{eq:ev2}
\end{eqnarray}
where ${\cal D}_{\veczeta}[z]$ is the
probability density function of the cross-talk noise
$z_{i}(\bs)$,
\begin{eqnarray*}
&&{\cal D}_{\veczeta}[z]
\equiv
2^{c}\left<\frac{1}{N}\sum_{i}
              \delta(z-z_{i}(\bs))
              \delta_{\veczeta \vecxi_{i}} \right
              >_{\vecm, r; t},\\
&& \int dz [ {\cal D}_{\veczeta}[z]]_{\veczeta}=1,
\end{eqnarray*}
where $\vecxi_i= ^t(\xi _i ^1,\cdots, \xi_i^{c})$.  
 $\bra \Phi \ket _{\vecm, r; t}$ denotes the subshell average
\[
 \bra \Phi \ket _{\vecm, r; t}
=\frac{\sum_{\bs} p_t(\bs)\delta(\vecm - \vecm(\bs))
\delta(r -r(\bs))\Phi(\bs)}
{\sum_{\bs} p_t(\bs)\delta(\vecm - \vecm(\bs))\delta(r -r(\bs))}.
\]
Here, to obtain closed equations, we  assume the following:
\begin{enumerate}
\item The time evolutions of the order parameters $(\vecm, r)$ have
the self-averaging property with respect to 
the average over the variables $\xi _i ^\mu$.
\item The probability distribution of the micro variable $\bs$
is independent of time and is 
uniform in the subshell where the macro variables
$(\vecm, r)$ have the same values.
\end{enumerate}
Then, we obtain the closed equations for  $(\vecm, r)$
 using the replica method under the RS ansatz,
\begin{eqnarray}
 \frac{d}{dt} \vecm &  = &  -\vecm + [ \veczeta \int Dx \int Dy 
 \{1-\tanh(^t \veczeta  \vecmu +y
\sqrt{\frac{\Delta}{2 \epsilon \rho}} \lambda +
\frac{\lambda ^2}{\sqrt{2 \epsilon} \rho} x )\} \nonumber \\
&& \times
 \tanh\beta( ^t\veczeta \matA \vecm +U^-)
] _{\veczeta},\\
 \frac{1}{2} \frac{d}{dt} r&=&  -r+1 + [ \int Dx \int Dy
\frac{1}{ \alpha} \{ 1-\tanh( ^t \veczeta  \vecmu +y
\sqrt{\frac{\Delta}{2 \epsilon \rho}} \lambda +
\frac{\lambda ^2}{\sqrt{2 \epsilon} \rho} x )\} \nonumber\\
&& \times U^-
 \tanh\beta(^t \veczeta \matA \vecm +U^-)
] _{\veczeta},
\end{eqnarray}
where $\epsilon=\frac{\alpha r}{2}, \Delta=\frac{\alpha \rho
(1-q)}{1-\rho (1-q)}$,
 $ U^{-}(x)=\sqrt{2\epsilon}x - \Delta$. 
When $(\vecm, r)$ are given, the parameters $\lambda, \rho, q$ and $ \bmu$
are determined using the following equations.
\begin{eqnarray}
\lambda &=& \frac{\rho \sqrt{\alpha q}}{1-\rho (1-q)},\\
r& = & \frac{1-\rho(1-q)^2}{[1-\rho (1-q)]^2}, \label{eq:rho}\\
q&=& [ \int Dw \tanh^2(^t \veczeta \vecmu + \lambda w)
] _{\veczeta}, \\
\vecm &=& [ \int Dw \veczeta \tanh(^t \veczeta \vecmu + \lambda w)
] _{\veczeta}.
\end{eqnarray}
From eq. (\ref{eq:rho}), $\rho$ is given by
\begin{equation}
\rho=\frac{1}{2r(1-q)}[2r - 1 + q - \sqrt{(1-q)^2+4rq}]\ge 0.
\end{equation}
In particular, the evolution equations for $T=0$ is given by
\begin{eqnarray}
\frac{d}{dt} \vecm
 &  =&  - \vecm + [ \veczeta  \int Dy \{
 \int_{0}^{\infty} \frac{du}{\sqrt{2\pi}} e^{-\frac{1}{2}(u+x_-)^2}
 \left(1 -\tanh( \Xi(u+x_-, y, \veczeta))\right) \nonumber \\
&&  -
  \int_{0}^{\infty}\frac{du}{\sqrt{2\pi}} e^{-\frac{1}{2}(u-x_-)^2}
 \left(1 -\tanh( \Xi(-(u-x_-),y, \veczeta))\right)
 \} ] _{\veczeta} , \label{eq:ev1T0}\\
\frac{1}{2} \frac{d}{dt} r &  =& 
  - r+1 + \frac{1}{\alpha}
[  \int Dy  \{
 \int_{0}^{\infty} \frac{du}{\sqrt{2\pi}} e^{-\frac{1}{2}(u+x_-)^2}
 \left(1 -\tanh( \Xi(u+x_-, y, \veczeta))\right)U^-(u+x_-)\nonumber\\
&& -
  \int_{0}^{\infty}\frac{du}{\sqrt{2\pi}} e^{-\frac{1}{2}(u-x_-)^2}
 \left(1 -\tanh ( \Xi(-u+x_-, y, \veczeta))\right) U^-(-u+x_-)\}
 ] _{\veczeta},\label{eq:ev2T0}
\end{eqnarray}
where $\Xi(x, y, \veczeta)   \equiv   \veczeta  \vecmu +y
\sqrt{\frac{\Delta}{2 \epsilon \rho}} \lambda +
\frac{\lambda ^2}{\sqrt{2 \epsilon} \rho} x$    and
$ x_{-}\equiv -\frac{1}{ \sqrt{2\epsilon}}
( ^t \veczeta \matA \vecm  - \Delta)$.

In the next subsection, we give the results for $T=0$
by the numerical calculations of the S.P.E., 
 and also the results of   the evolution equations
 together with the results by numerical simultions.

\subsection{Results}
\noindent
{\bf The equilibrium states}

We solved the S.P.E. (\ref{eq:spe1})-(\ref{eq:spe3}) numerically 
for several values of parameters $a$ and $\alpha$ and
found the coexistence region of  correlated attractors,
Hopfield attractors, and mixed states.   
In fig. 5, we show an example of the phase diagrams.
When $\alpha$ is decreased, at $\alpha \simeq 0.0183$, a pair of
correlated attractors appear.  One of the correlated attractors 
continues to exist at least until $\alpha \simeq 0.0014$.
This attractor is not shown in fig. 5 because it is unstable,
that is, this is not an attractor but a repeller.
Another correlated attractor,
 which is drawn in the figure, is stable and exists until 
$\alpha \simeq 0.0049$.
At $\alpha \simeq 0.0049$, this correlated attractor disappears
and then a mixed state with three patterns appears and continues
to exist until $\alpha = 0$.
Further, a mixed state with thirteen patterns exist for
$\alpha \in (0, 0.3119)$.
The Hopfield attractor exists only for $\alpha \in (0, 0.013)$.
Thus, for $\alpha \in (0.0049, 0.013)$, the Hopfield attractor,
a correlated attractor, and a mixed state with thirteen patterns coexist.\\

\noindent
{\bf Dynamics}

We studied the following two cases using deterministic dynamics.

{\bf Case 1} A Hopfield attractor 
and a correlated attractor coexist.

{\bf Case 2} A correlated attractor, but no Hopfield attractor.

In both cases, we performed numerical integrations 
of the evolution equations (\ref{eq:ev1T0}) and (\ref{eq:ev2T0})
and numerical simulations of eq. (\ref{eq:amit1}).
In the numerical integration, we took 
$\vecm(0)=(m_0, 0,\cdots,0)$ and $r(0)=1$ as an initial condition.
Since the integration of the evolution equations required
 a lot of computation time,
we used the Euler method as the integration scheme.

In the numerical simulations,  an initial value of $s_i$
was determined according to the probability distribution (\ref{eq:prob}).
In both numerical integrations and numerical simulations, $m_0$
was taken as $m_0=0.1,0.2,\cdots,0.9$.

{\bf Case 1}

As an example, we set $a=0.35, \alpha=0.01$.
 In fig. 6, we show the result of DRT and 
of numerical simulations.
As the figure shows, the boundary between  the basin of 
attraction for the Hopfiled attractor and that for the
 correlated attractor is around $m_1(0) = 0.5$.
Except for the boundary of the basin of attraction,
the results by two methods agree fairly well.

{\bf Case 2}

As an example, we set $a=0.35$ and $ \alpha=0.015$.
As is seen from fig. 7, the trajectories initially
approach  the state where the Hopfield attractor existed but
finally tend toward the correlated attractor
in both the numerical integrations and numerical simulations
except for $m_0 =0.9$ in the numerical integration.
In the case of $m_0=0.9$ in the numerical integration,
our numerical scheme to solve the saddle point
equations did not generate a solution at some time.
We suggest
that this is due to the precision of the numerical calculations.\\
The tendency was that trajectories obtained by numerical integrations
tend toward the correlated attractor faster than those obtained by 
numerical simulations.

\section{Summary and Discussion}

We studied the statics and dynamics of the  model introduced by 
Griniasty et al. in the finite and extensive loading cases.
In both cases, we qualitatively obtained similar results.
First, we summarize the results common to both cases.

As for the statics, we  solved the saddle point equations
numerically and obtained equilibrium states.
We found regions of parameters where the Hopfield attractor
and a correlated attractor coexist.
It had been previouly believed that 
correlated attractors exist only for $a>0.5$.
However, we found  that the correlated attractors exist
not only for $a>0.5$  but also for $a<0.5$. 

As for the dynamics,  we studied the following two cases 
by  numerical integrations of the evolution equations and
by the numerical simulations, 
deriving the evolution equations by dynamical replica theory
in the extensive loading case.

First, we studied the transient behaviour 
when a Hopfield attractor and a correlated attractor coexist.
We found that 
when we set initial conditions at $\vecm (0)=(m_0, 0,\cdots,0)$
( and $r(0)=1$ for the extensive loading case),
there exists some critical initial overlap $m_0^c$,
which is the boundary between the basin of attraction for the
Hopfield attractor and that for the correlated attractor.
Next, we studied the transient behaviour 
when a Hopfield attractor does not exist but a correlated attractor 
does exist.
Then, we found that  trajectories initially approach 
the state where the Hopfield attractor existed but finally
they tend toward the correlated attractor.

Next, we compare the results obtained in the
finite loading case and the results obtained 
in the extensive loading case.\\
\noindent
{\bf Equilibrium state}\par
In both cases, the Hopfield attractor, 
the mixed state, and the correlated attractor 
exist, and
there are parameter regions where the
 Hopfield attractor, the mixed state, and the correlated attractors
coexist.\\
\noindent
{\bf Dynamics}\\
\noindent
{\bf 1. When several attractors coexist}\par
In the finite loading case ($a=0.4, T=0.04$),
the boundary between  the basin of attraction 
for the Hopfiled attractor and that for the correlated attractor is
$0.15< m_1 ^c< 0.16$ in the numerical integrations and
$0.16< m_1 ^c< 0.17$ in the numerical simulations.
On the other hand, in the
extensive loading case($a=0.35, \alpha=0.01, T=0$),
the boundary between  the basin of attraction 
for the Hopfiled attractor and that for the correlated attractor is
$0.4< m_1 ^c < 0.5$ in both the numerical integrations and
the numerical simulations.
In the finite and extensive loading cases,
the results of numerical integrations 
and the numerical simulations agree fairly
well.
One reason for the difference between the finite and extensive loading cases
might be due to the difference in parameters.\\
\noindent
{\bf 2. When  a Hopfield attractor does not exist 
but a correlated attractor does exist}\par
In the finite loading case, we used  $m_1(0)=0.1,0.2,\cdots,1.0$ 
as the initial conditions.
Then the trajectories finally tended toward the correlated attractor in both the
numerical integrations and numerical simulations.
On the other hand, for the
extensive loading case ($a=0.35, \alpha=0.015, T=0$),
we used $m_1(0)=0.1,0.2,\cdots, 0.9$ and $r(0)=1$
as initial conditions. 
Then, trajectories finally tended toward the correlated attractor in both the
numerical integrations and numerical simulations, except
for  $m_1(0)=0.9$ in the numerical integration
for which it seems that the precision of the numerical calculation
is not good enough.  Except for the last case,
the results of numerical integrations
and numerical simulations  agree fairly
well in both the finite and extensive loading cases.\\

Therefore, we demonstrate that the results similar to those 
in the finite loading case have been obtained in the extensive loading
case.  \par
It might be considered that if we could construct a physiological
experiment to observe the transient behaviour of a firing pattern
of a monkey, we would be able to judge whether  an
attractor neural network exists in the brain of the monkey.
As Miyashita and Chang reported,
the neurons in the AVT are highly selective towards a few of the 100
colored fractal patterns.
This implies that the firing rate of neurons in the AVT is very low,
while the firing rate is set to 50$\%$ in the present model.
Thus, in the future it will be necessary 
to develop a model with a low firing rate,\cite{Okada1996}
 so that we can compare results of that model with
physiological experimental results.

\section*{Acknowledgements}
We are grateful to A. C. C. Coolen 
 for valuable discussions and suggestions of this work.
This work was partially supported by Grant-in-Aid for Scientific
Research on Priority Areas No. 14084212 and Grant-in-Aid for
Scientific Research (C) No. 14580438.


\newpage

\begin{figure}[t]
\vspace*{1mm} \setlength{\unitlength}{1.04mm}
\begin{picture}(150,90)
 \put(5,20){\epsfxsize=90\unitlength\epsfbox{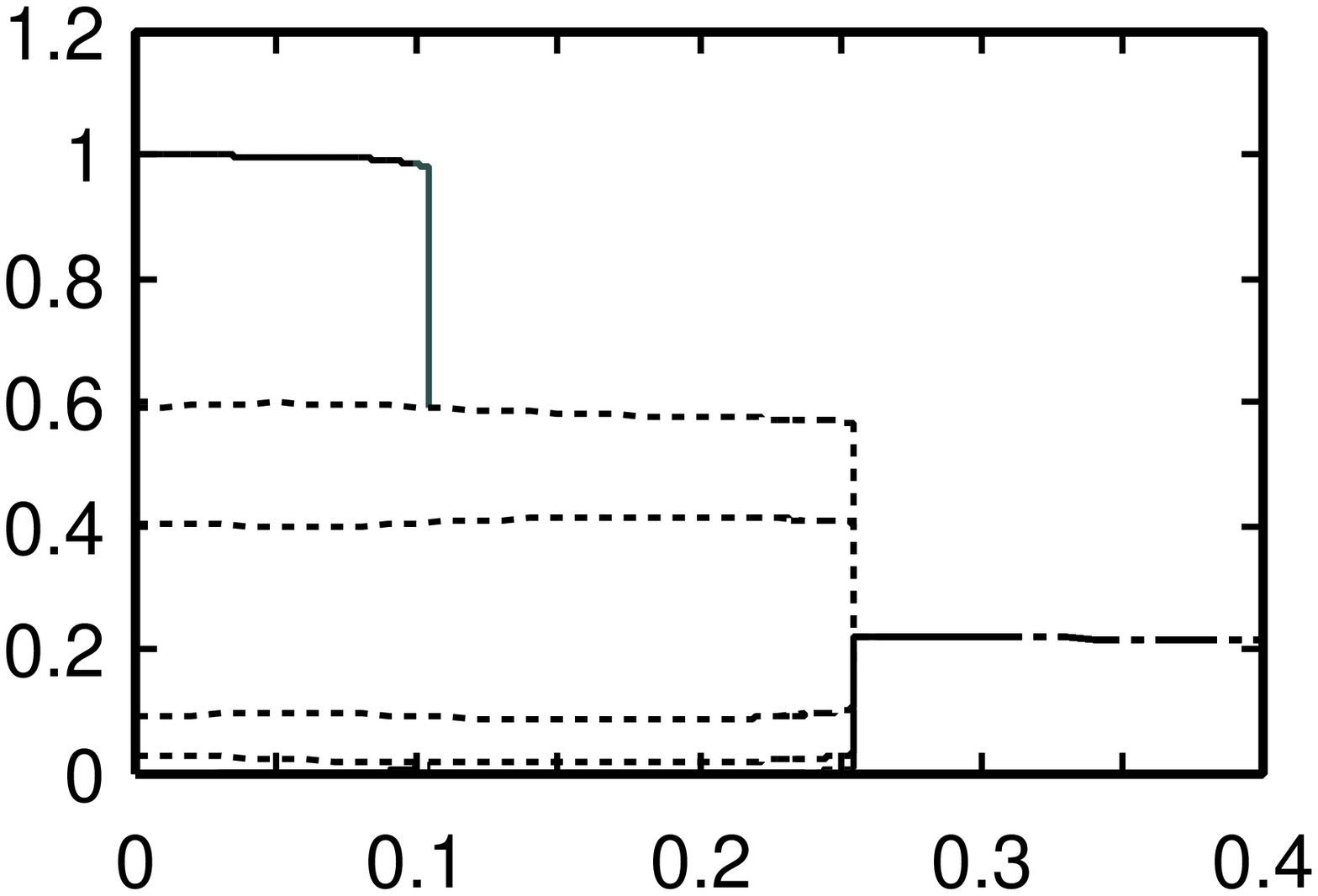}}
\put(-1,55){$m_{\mu}$} \put(50,15){$T$}
\end{picture}
\vspace*{-13mm}
 \caption{
Equilibrium states in finite loading case.
$a=0.4$ and $c=13$. 
The abscissa is $T$ and the ordinate is $m_{\mu}(\mu =1, \cdots, 7)$.
Solid curve: Hopfield attractor. Dashed curves: correlated attractor.
Dot-dash-curve: mixed state of thirteen patterns.
}
\end{figure}

\begin{figure}[htb]
\vspace*{1mm} \setlength{\unitlength}{1.04mm}
\begin{picture}(100,200)
\put(7,120){\epsfxsize=100\unitlength\epsfbox{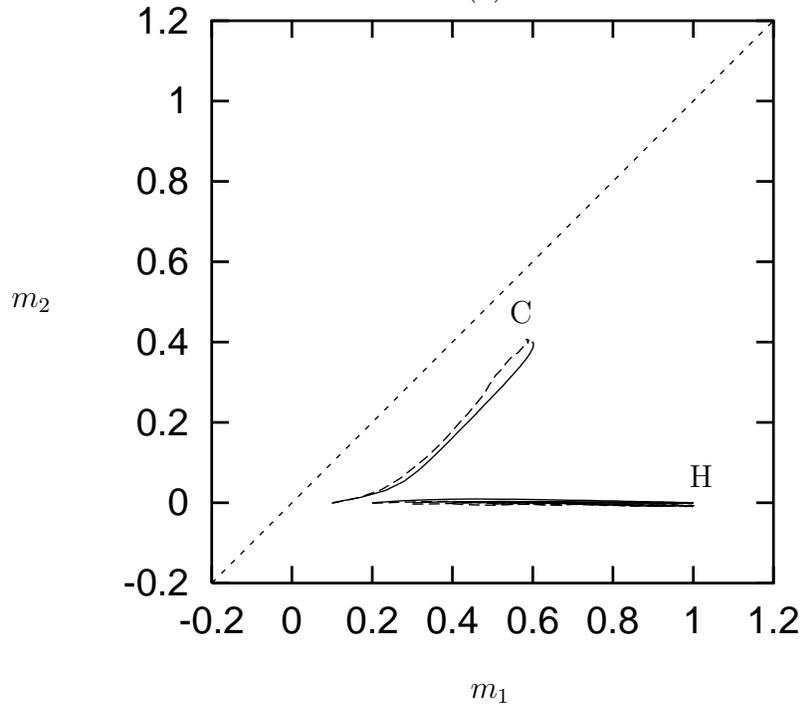}}
\put(65,163){C} \put(88,142){H}
 \put(58,205){ (a)}
 \put(0,165){ $m_2$}   \put(60,115){$m_1$}
 \put(7,20){\epsfxsize=100\unitlength\epsfbox{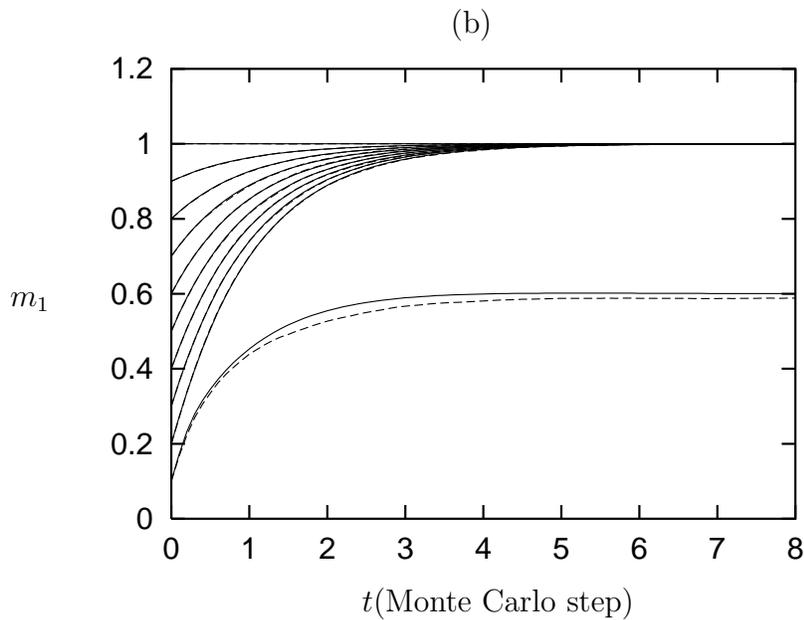}}
 \put(55,90){ (b)}
\put(0,55){$m_1$} \put(45,16){$t$(Monte Carlo step)}
\end{picture}
\vspace*{-13mm}
 \caption{
Results of numerical integrations and numerical
 simulations for the finite loading case.
$a=0.4, T=0.04$ and $ c=13$.
Initial condition is $\vecm=(m_0, 0,\cdots,0)$ and
$m_0=0.1, 0.2, \cdots, 1.0$.
 Dashed curves: simulation for $N=60,000$. Solid curves: theory.
(a) Trajectory in  space of $(m_1, m_2)$.
C: correlated attractor. H: Hopfield attractor.
(b)Time evolution of $m_1$.
}
\end{figure}

\begin{figure}[htb]
\vspace*{1mm} \setlength{\unitlength}{1.04mm}
\begin{picture}(100,200)
\put(7,120){\epsfxsize=100\unitlength\epsfbox{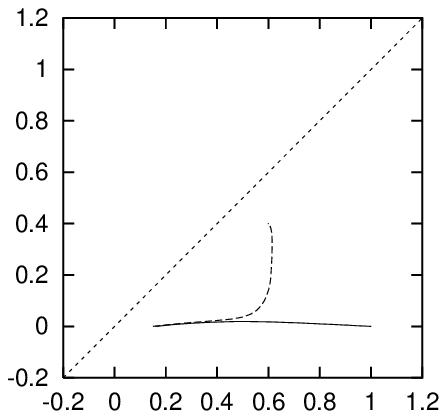}}
\put(65,163){C} \put(88,142){H}
 \put(58,205){ (a)}
 \put(0,165){ $m_2$}   \put(60,115){$m_1$}
 \put(7,20){\epsfxsize=100\unitlength\epsfbox{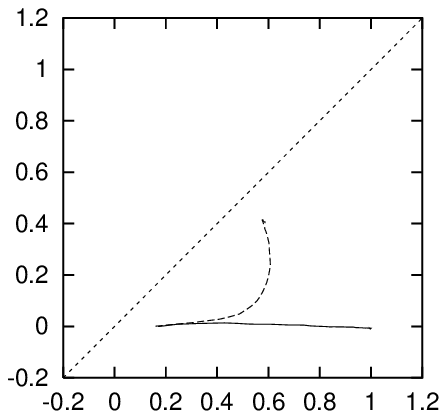}}
\put(65,63){C} \put(88,42){H}
 \put(58,105){ (b)}
\put(0,65){$m_2$} \put(60,15){$m_1$}
\end{picture}
\vspace*{-13mm}
 \caption{
Results of numerical integrations (a) and numerical
 simulations (b)  for finite loading case.
 $a=0.4$, $T=0.04$ and $c=13$. Initial condition is $\vecm=(m_0, 0,\cdots,0)$.
(a) Dashed curve: $m_0$ = 0.15. 
Trajectory tends toward  correlated attractor.
Solid curve: $m_0$ = 0.16. Trajectory tends toward Hopfield attractor.
(b)
 $N=60,000$.
Dashed curve: $m_0$ = 0.16.  Trajectory  tends toward  correlated attractor.
Solid curve: $m_0$ = 0.17.  Trajectory tends toward Hopfield attractor.
}
\end{figure}

\begin{figure}[htb]
\vspace*{1mm} \setlength{\unitlength}{1.04mm}
\begin{picture}(100,200)
\put(7,120){\epsfxsize=100\unitlength\epsfbox{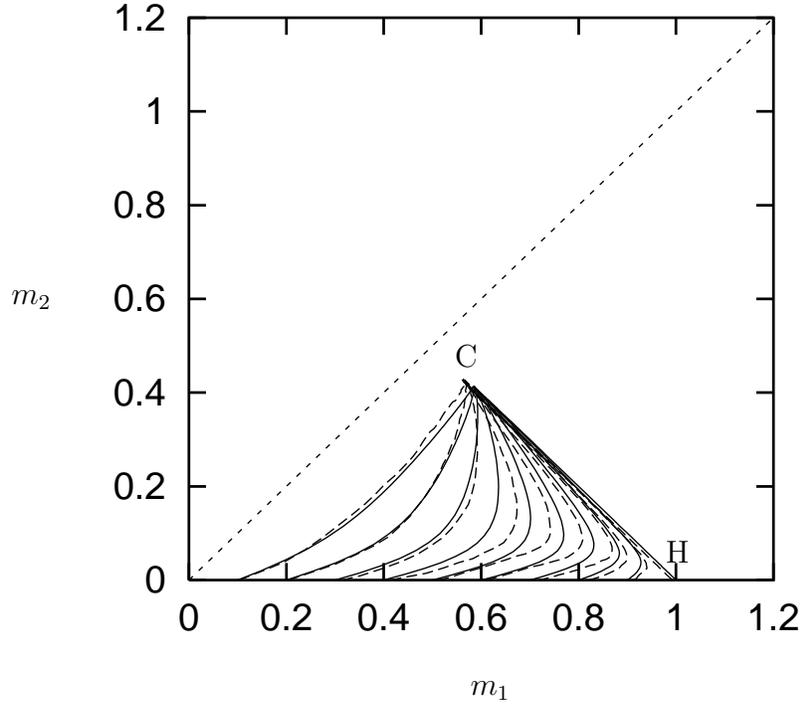}}
\put(58,157){C} \put(85,132){H}
 \put(58,205){ (a)}
 \put(0,165){ $m_2$}   \put(60,115){$m_1$}
 \put(7,20){\epsfxsize=100\unitlength\epsfbox{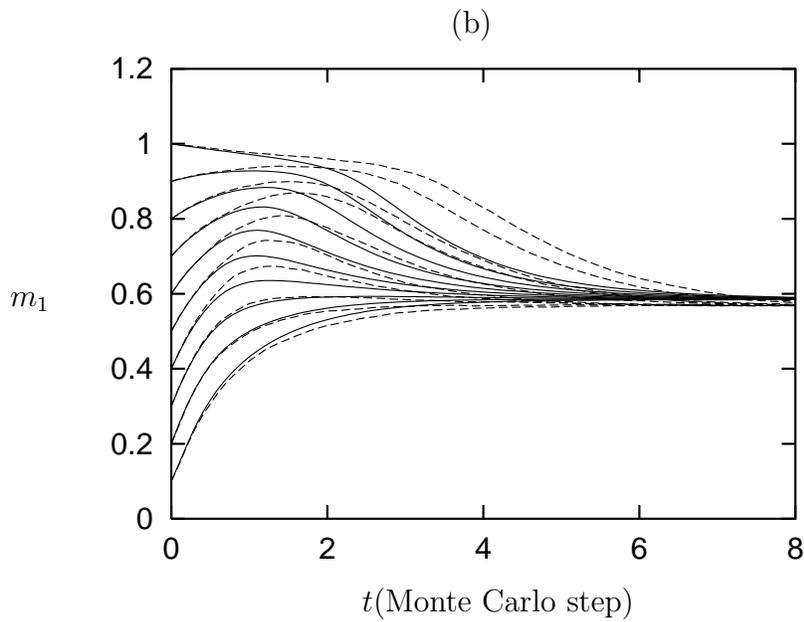}}
 \put(55,90){ (b)}
\put(0,55){$m_1$} \put(45,16){$t$(Monte Carlo step)}
\end{picture}
\vspace*{-13mm}
 \caption{
Results of numerical integrations and numerical
 simulations for finite loading case.
$a=0.4, T=0.15$ and $ c=13$.
 Dashed curves: simulation for $N=60,000$. Solid curves: theory.
 (a) Trajectory in space of $(m_1, m_2)$.
C: correlated attractor. H:  point where Hopfield attractor existed.
(b) Time evolution of $m_1$.
}
\end{figure}

\begin{figure}[t]
\vspace*{1mm} \setlength{\unitlength}{1.04mm}
\begin{picture}(100,80)
\put(10,5){\epsfxsize=100\unitlength\epsfbox{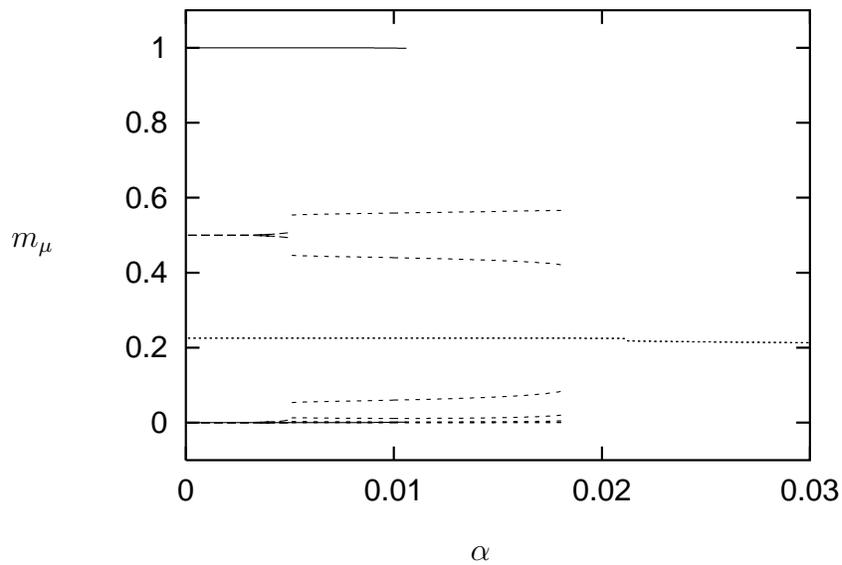}}
 \put(0,40){ $m_{\mu}$}   \put(60,0){$\alpha$}
\end{picture}
 \caption{
Equilibrium states for $a=0.35$ and $T=0$
 for case of extensive loading. The abscissa is $\alpha$
and the ordinate is $m_{\mu}( \mu =1, \cdots, 7)$.
Solid curve: Hopfield attractor. Dashed curves: correlated attractor.
Long dashed curves: mixed state of three patterns.
Dotted curve: mixed state of thirteen patterns.
}
\end{figure}

\begin{figure}[htb]
\vspace*{1mm} \setlength{\unitlength}{1.04mm}
\begin{picture}(100,200)
\put(7,120){\epsfxsize=100\unitlength\epsfbox{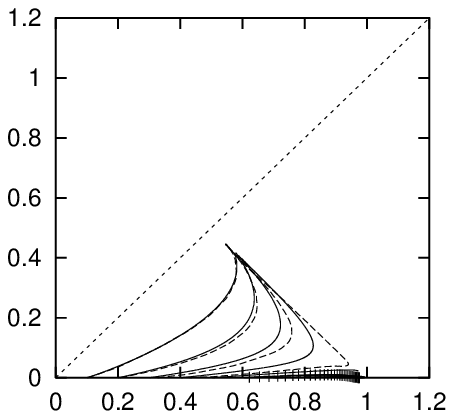}}
\put(54,155){C} \put(85,132){H}
 \put(58,205){ (a)}
 \put(0,165){ $m_2$}   \put(60,115){$m_1$}
 \put(7,20){\epsfxsize=100\unitlength\epsfbox{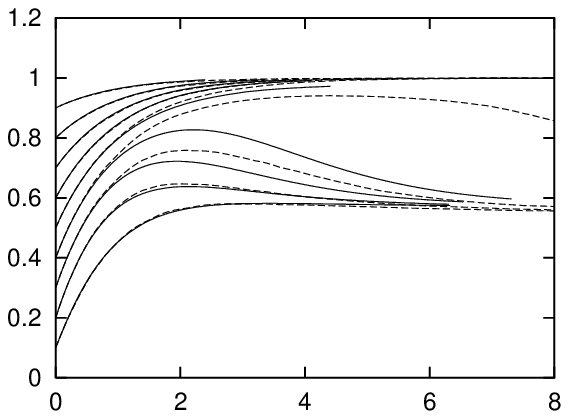}}
 \put(55,90){ (b)}
\put(0,55){$m_1$} \put(45,16){$t$(Monte Carlo step)}
\end{picture}
\vspace*{-13mm}
 \caption{
Results of numerical integrations and numerical
 simulations for extensive loading case.
$a=0.35, \alpha=0.01, T=0$ and $c=13$.
 Dashed curves: simulation for $N=60,000$. Solid curves: DRT.
(a) Trajectory in  space of $(m_1, m_2)$.
C: correlated attractor. H: Hopfield attractor.
(b)Time evolution of $m_1$.
}
\end{figure}

\begin{figure}[htb]
\vspace*{1mm} \setlength{\unitlength}{1.04mm}
\begin{picture}(100,200)
\put(7,120){\epsfxsize=100\unitlength\epsfbox{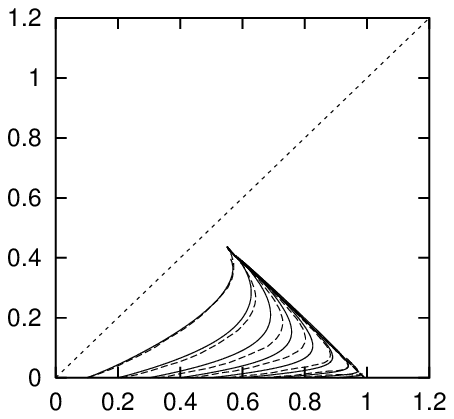}}
\put(54,155){C} \put(85,132){H}
 \put(58,205){ (a)}
 \put(0,165){ $m_2$}   \put(60,115){$m_1$}
 \put(7,20){\epsfxsize=100\unitlength\epsfbox{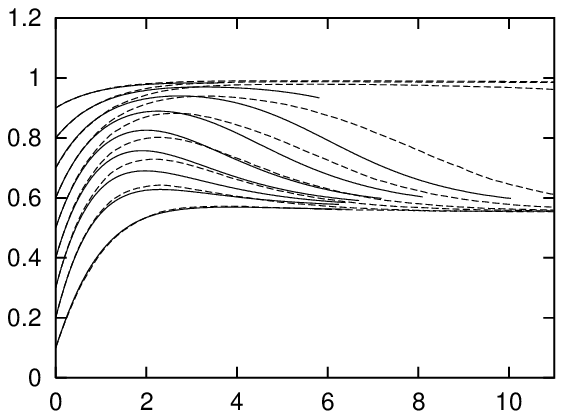}}
 \put(55,90){ (b)}
\put(0,55){$m_1$} \put(45,16){$t$(Monte Carlo step)}
\end{picture}
\vspace*{-13mm}
 \caption{
Results of numerical integrations and numerical
 simulations for the extensive loading case.
 $a=0.35, \alpha=0.015, T=0$ and $c=13$.
 Dashed curves: simulation for $N=60,000$. Solid curves: DRT.
(a) Trajectory in space of $(m_1, m_2)$.
C: correlated attractor. H: point where  Hopfield attractor existed.
(b)Time evolution of $m_1$.
}
\end{figure}


\end{document}